\documentclass[prl,aps,twocolumn,floatfix]{revtex4}
\pdfoutput=1
\usepackage{graphicx}
\usepackage[small,centerlast]{caption}
\usepackage{subfig}

\begin{document}
\title{Dual Jets from Binary Black Holes} 
\author
{Carlos Palenzuela,$^{1,2}$ Luis Lehner,$^{3,4,5}$ and Steven L. Liebling$^{6}$}
\affiliation{
{$^{1}$Canadian Institute for Theoretical Astrophysics, Toronto, Ontario M5S 3H8, Canada}\\
{$^{2}$Dept. of Physics \& Astronomy, Louisiana State University, Baton Rouge, LA 70803, USA}\\
{$^{3}$Perimeter Institute for Theoretical Physics, Waterloo, Ontario N2L 2Y5, Canada}\\
{$^{4}$Department of Physics, University of Guelph,
Guelph, Ontario N1G 2W1, Canada}\\
{$^{5}$Canadian Institute For Advanced Research (CIFAR),  Cosmology and Gravity Program, Canada}\\
{$^{6}$Department of Physics, Long Island University, New York, USA}}

\begin{abstract}
Supermassive black holes are found at the centers of most galaxies and
their inspiral is a natural outcome when galaxies merge. The inspiral of these
systems is of utmost astrophysical importance as prodigious producers
of gravitational waves and in their possible role in energetic electromagnetic
events. We study such binary black hole coalescence under the influence of
an external magnetic field produced by the expected circumbinary disk surrounding them.
Solving the Einstein equations to describe the
spacetime and using the force-free approach for the electromagnetic fields and the
tenuous plasma, we present numerical evidence for possible jets driven by these systems.
Extending the process described by Blandford and Znajek for a single spinning black hole,
the picture that emerges suggests the electromagnetic field extracts energy from
the orbiting black holes, which ultimately merge and settle into the standard Blandford-Znajek
scenario. Emissions along dual and single jets
would be expected that could be observable to large
distances.
\end{abstract}

\maketitle 

\noindent{\bf{\em Introduction:}}
Among the most spectacular of astrophysical events are observations of
tremendous amounts of energy careening out along highly
collimated jets from a localized central region.  Within this central
region is the engine behind the jets, generally believed to be a spinning
black hole which helps to convert
  binding and rotational energy 
into kinetic and thermal energy of the surrounding plasma (see e.g.~\cite{2002luml.conf..381B}).
This basic picture is widely accepted, but a detailed understanding
of these systems remains elusive in part due to the inability of
detecting clean electromagnetic signals from the depths of the central engine. 
Such understanding is particularly needed to understand phenomena such as
gamma ray bursts and AGNs, which radiate a tremendous amount of energy and, in some cases,
are thought to play a fundamental role in the formation and evolution of massive 
galaxies~\cite{2006MNRAS.370..645B}. 

In addition to intense efforts in various electromagnetic bands,
observations in the gravitational wave band will soon be added. 
Because gravitational waves propagate essentially unscattered through the universe and are
most strongly produced in highly dynamical regions of strong spacetime curvature, their observation 
promises to open up a new complementary window to the universe, 
allowing for multi-messenger astronomy~\cite{1997RvMA...10....1T,2009LRR....12....2S}.
Astrophysical systems expected to be bright in both
gravitational and electromagnetic spectrums are therefore attracting intense attention.

A particularly interesting scenario involves the merger of a supermassive binary black system which
could be observable with LISA to redshifts up to $z=5$ and beyond~\cite{Vecchio:2003tn}. 
Such a system is expected as a natural consequence of galaxy mergers, whose individual black
holes eventually settle into an orbit sufficiently
tight that the binary's subsequent dynamics is governed by the gravitational radiation timescale and 
 disconnected from the properties of the disk~\cite{Milosavljevic:2004cg}.
As they orbit, they emit gravitational waves which carry off both energy and angular momentum, driving
the black holes to merger  producing bright gravitational waves.
The expected gravitational wave signals from such a system, their impact on gravitational searches and
their exploitation for physics mining through data analysis 
are now well understood thanks to concentrated efforts on the analytical 
and numerical fronts (see e.g.~\cite{Aylott:2009ya} and references cited therein). Such efforts
have studied the dynamics of orbiting binary black holes in vacuum, which is a good approximation
as the black hole's inertia is orders of magnitude larger than any other one.

A logical next step is
to comprehend possible electromagnetic counterparts which would
allow for studying key systems via both bands. Such a possibility is quite strong in the context
of supermassive binary black hole mergers, because in the merger process the black holes will 
interact with at least: (i) a circumbinary 
accretion disk~\cite{Milosavljevic:2004cg}, (ii)
remnant gas in between the black holes~\cite{2005ApJ...634..921A,2008ApJ...672...83M}, 
(iii) a magnetosphere~\cite{Blandford:1977ds} and give rise to possible 
electromagnetic emissions (e.g.~\cite{kickdisk1,kickdisk2,massloss,kickdisk,Corrales:2009nv,Rossi:2009MNRAS,Zanotti:2010xs};
~\cite{Chang:2009rx,Bode:2009mt,Krolik:2009hx};~\cite{Palenzuela:2009yr,Palenzuela:2009hx,Mosta:2009rr}).
As discussed in ~\cite{Milosavljevic:2004cg}, as the binary tightens, a common circumbinary
disk will be formed within which the black holes will eventually merge.  Such a disk will 
typically be magnetized, anchoring a magnetic field which will permeate
the region inside the disk containing the black holes.
Once the black holes merge, the generic outcome will be a spinning black hole within the magnetic
field anchored by the disk. It is precisely this system studied in the seminal work
of Blandford and Znajek~\cite{Blandford:1977ds} and a large body of subsequent works.
In this model, the spinning black hole immersed in an external magnetic field accelerates 
stray electrons to energies sufficient to produce cascading pair-production which supplies
the environment with a plasma~\cite{2002luml.conf..381B}. 
Due to the negligible fluid inertia, the bulk of
such plasma moves freely without experiencing any Lorentz force,
and so the system can be cleanly studied with the so-called
force-free approximation~\cite{1969ApJ...157..869G,Blandford:1977ds}.
This approximation, together with the assumption of 
stationarity and axisymmetry, allows one to determine that a net outward electromagnetic flux
is produced, which extracts rotational energy from the black hole. Furthermore, the electromagnetic
energy flux is highly collimated and eventually will be transferred into kinetic energy of
the plasma, accelerating charges and producing synchrotron radiation. Support for this
basic picture has been provided in recent years through numerical models~\cite{Komissarov:2004b,2004Sci...305..978S,Komissarov:2007rc,2008MNRAS.388..551T,Krolik:2009rn}.

The Blandford-Znajek model outlined above is not directly applicable to the very dynamical 
merger of two black holes, thus  the interaction of  electromagnetic fields and plasma and their
possible emissions in a such system has remained a mystery.
Our studies indicate that the dynamical behavior of the system: induces a collimation around
each intervening black hole --generating a toroidal magnetic field---;
amplifies the electromagnetic field strength (when comparing to the single black hole case),
and emits a strong electromagnetic burst associated with the merger. 
The collimated area can thus become a channel for accelerated particles which, in turn, 
can emit observable electromagnetic radiation, in particular of synchrotron type.
Furthermore the behavior of the EM fields, tightly tied to the binary's dynamics,
make them strong tracers of the spacetime and the collimation tubes can be seen as a cover
for the familiar ``pair of pants'' description of the black holes' event horizons~\cite{Matzner:1995ib}.\\

\noindent{\bf{\em Implementation details:}}
We study the black hole binary via numerical simulations which implement the (coupled) Einstein-Maxwell 
system of equations. Also coupled to the system is a low density plasma for which the inertia is
negligible compared with the electromagnetic stresses. The plasma essentially experiences no Lorentz force
and so the electromagnetic fields are well described by the force free condition
$\rho_{\rm EM} E + j_{\rm EM} \wedge B = 0$
(with $\rho_{\rm EM}$, $j_{\rm EM}$ the charge and current densities). The role of
the plasma is then to provide charge and current densities. The complete system is implemented 
using finite difference approximations within a computational infrastructure that 
provides distributed adaptive mesh refinement~\cite{had_webpage} with seven levels of refinement. 
Our computational domain is defined by 
$x^i \in[-1207 \, M_8 AU,1207 \, M_8 AU]$  
with a finest resolution of 
$\Delta = 0.4 \, M_8 AU$,
where $M$ is the total mass of the system in terms of the solar mass $M_{\odot}$ and
 $M_8 \equiv M/10^8 M_{\odot}$.
We adopt fourth order accurate spatial discretizations and third order Runge-Kutta integration in time,
while the singularity inside each black hole is excised from the computational domain. Since the region inside
the black hole is causally disconnected from the outside, this procedure does not affect the physics obtained
there.
More details of our techniques for implementing Einstein equations can be found in
e.g.~\cite{Calabrese:2003vx,Lehner:2005vc,Palenzuela:2006wp,Palenzuela:2009yr} 
and for our adoption of the force free condition we follow closely the approach presented 
in~\cite{Komissarov:2004b}.

The circumbinary disk is assumed to lie outside the computational domain and
so its details are unimportant. Such an assumption is strongly justified by the fact
that the viscosity in the disk is such that the disk cannot keep pace with the
shrinking orbit of the black holes, and as a result the disk ``freezes-in'' at
a radius typically considerably larger than our computational domain~\cite{Milosavljevic:2004cg}.

The disk's role is therefore to anchor the magnetic field which is 
incorporated as boundary and initial conditions describing a dipolar magnetic field
with strength $B_0$ in the region around the black holes. 
We adopt black holes in a quasi-circular orbit (which is
a reasonable approximation as gravitational wave emission circularizes the orbit).
We study late orbiting stages and merger, and so we concentrate on an initial separation 
that gives rise to over one orbit before the merger takes place. This scenario
allows us to study 
the transition of the binary from orbit to merger into a final black hole. Furthermore, we
adopt equal-mass ($M_{BH}=M/2$, with a radius $R_{BH}=10 M_8 AU$) non-spinning black holes to disentangle
orbitally-driven effects from those that would be induced by any individual spins (i.e. individual 
BZ effects that would be present around each spinning black hole). \\

\noindent{\bf{\em Results:}}
As mentioned, gravitational waves from this system, in which the plasma and electromagnetic inertia are 
orders of magnitude smaller than that of the black holes, are well understood by now (see~\cite{Aylott:2009ya} and
references cited there in)
and our results reproduce the expected behavior.
The electromagnetic field behavior on the other hand, is completely new and we concentrate 
on it here.
The global dynamical behavior is illustrated in Fig 1. The two black holes orbit and merge within
about an orbital period 
($\sim 13.6 \, M_8$ hours), after which the black final hole region radiates its
excess structure, settling into a Kerr black hole with spin 
$a \equiv J\,c/(G\,M^2) \simeq 0.67$. To study the system
in more detail, we monitor different quantities: The Newman-Penrose scalars $\Phi_2$ and
$\Psi_4$ (the square of the former and the integral of the latter describe the electromagnetic
and gravitational energy fluxes respectively); the Poynting flux; the function 
$\Omega_F \equiv F_{r\theta}/F_{r\phi}$ (which in axisymmetric, stationary scenarios
describes the angular velocity of magnetic field lines~\cite{Blandford:1977ds}) and the electric
and magnetic field topologies. To describe the behavior of the system, it is easier to concentrate
on the different stages identified in the binary black hole dynamics:

\begin{figure}
\centering
\subfloat[Part 1][$-11.0 \,M_8$ hrs]{\includegraphics[scale=0.25]{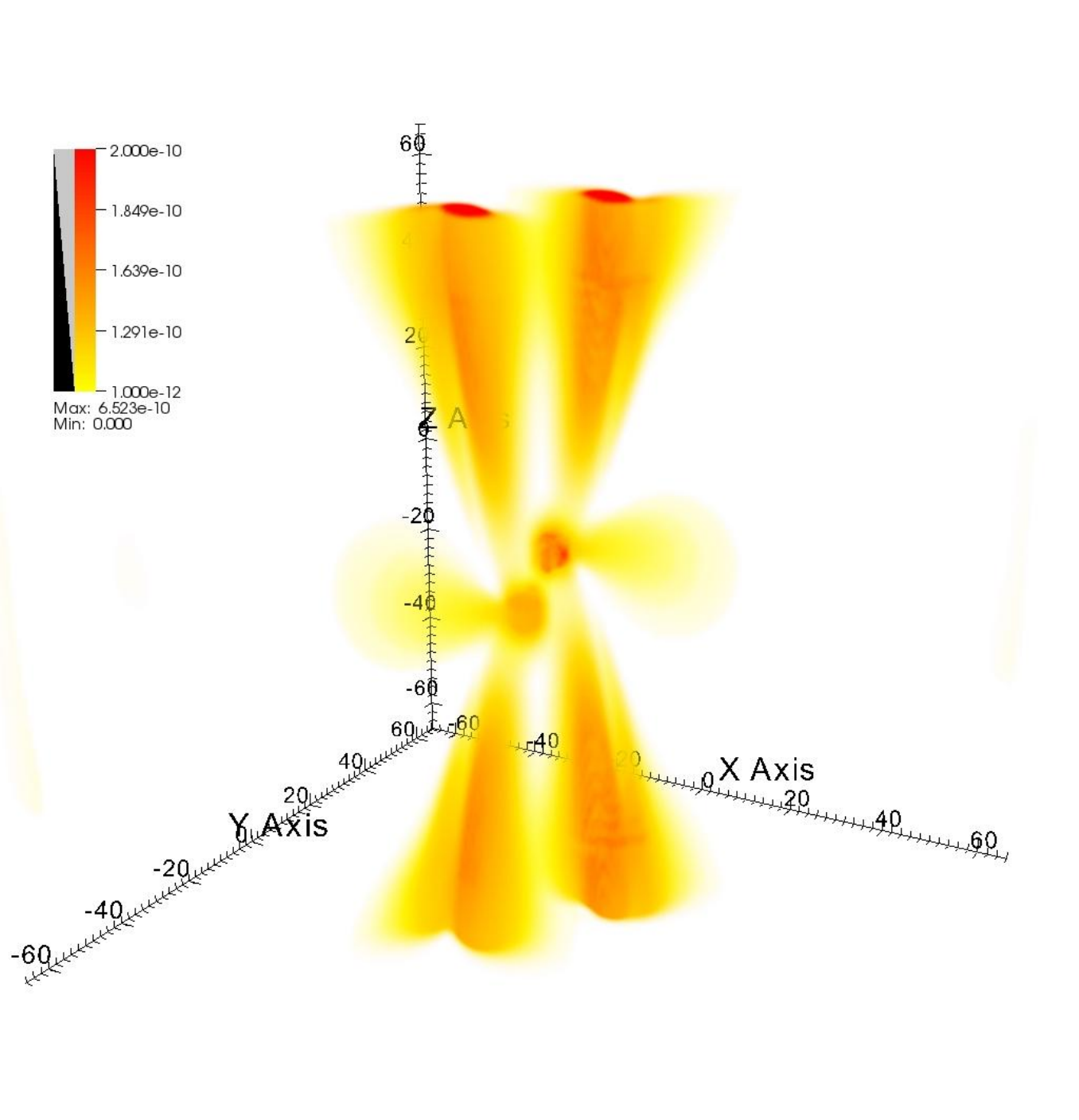} \label{fig:fourfigures-a}}
\subfloat[Part 2][$-3.0\, M_8$ hrs]{\includegraphics[scale=0.25]{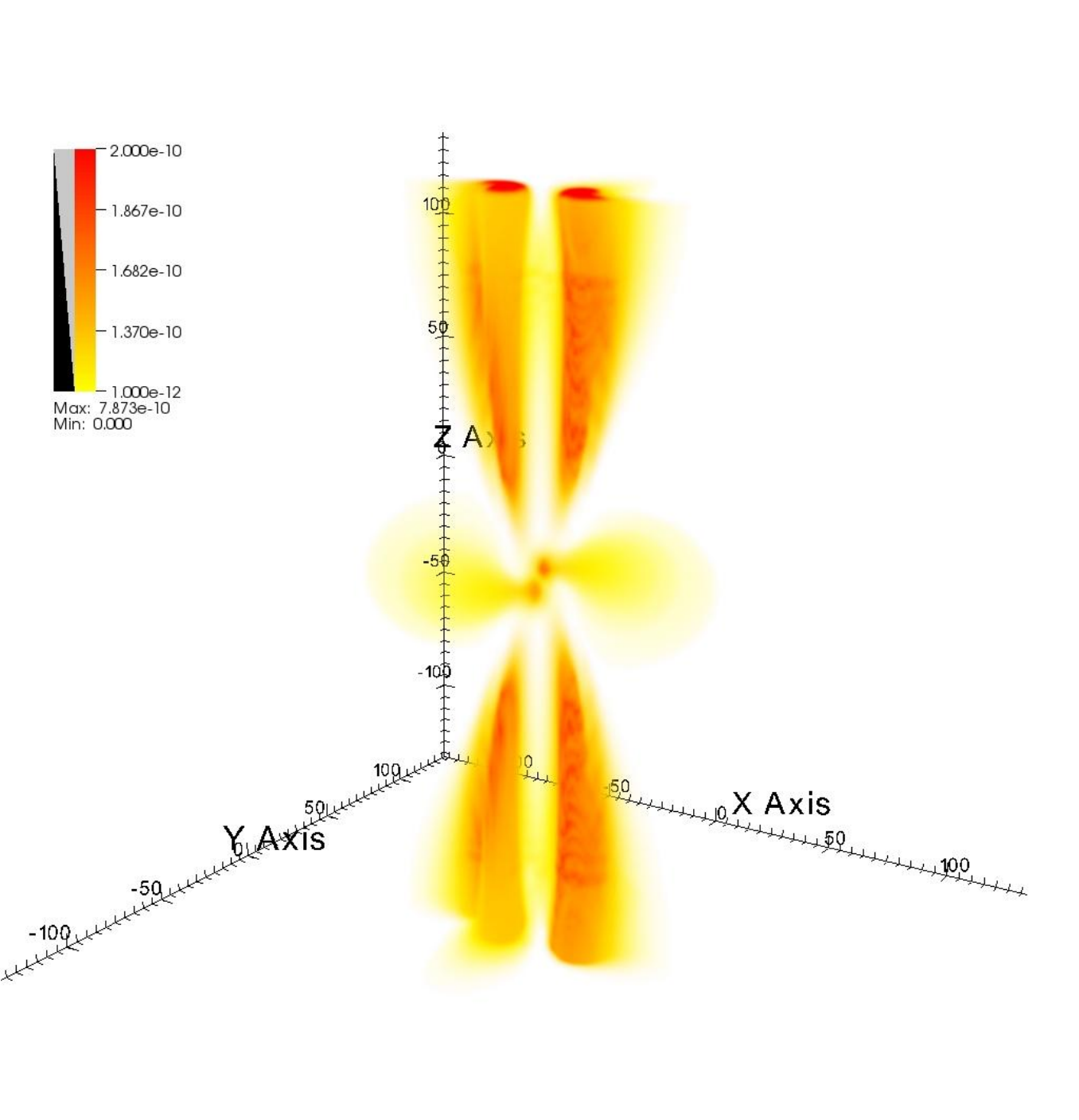} \label{fig:fourfigures-b}}\\
\subfloat[Part 3][$4.6 \, M_8$ hrs]{\includegraphics[scale=0.25]{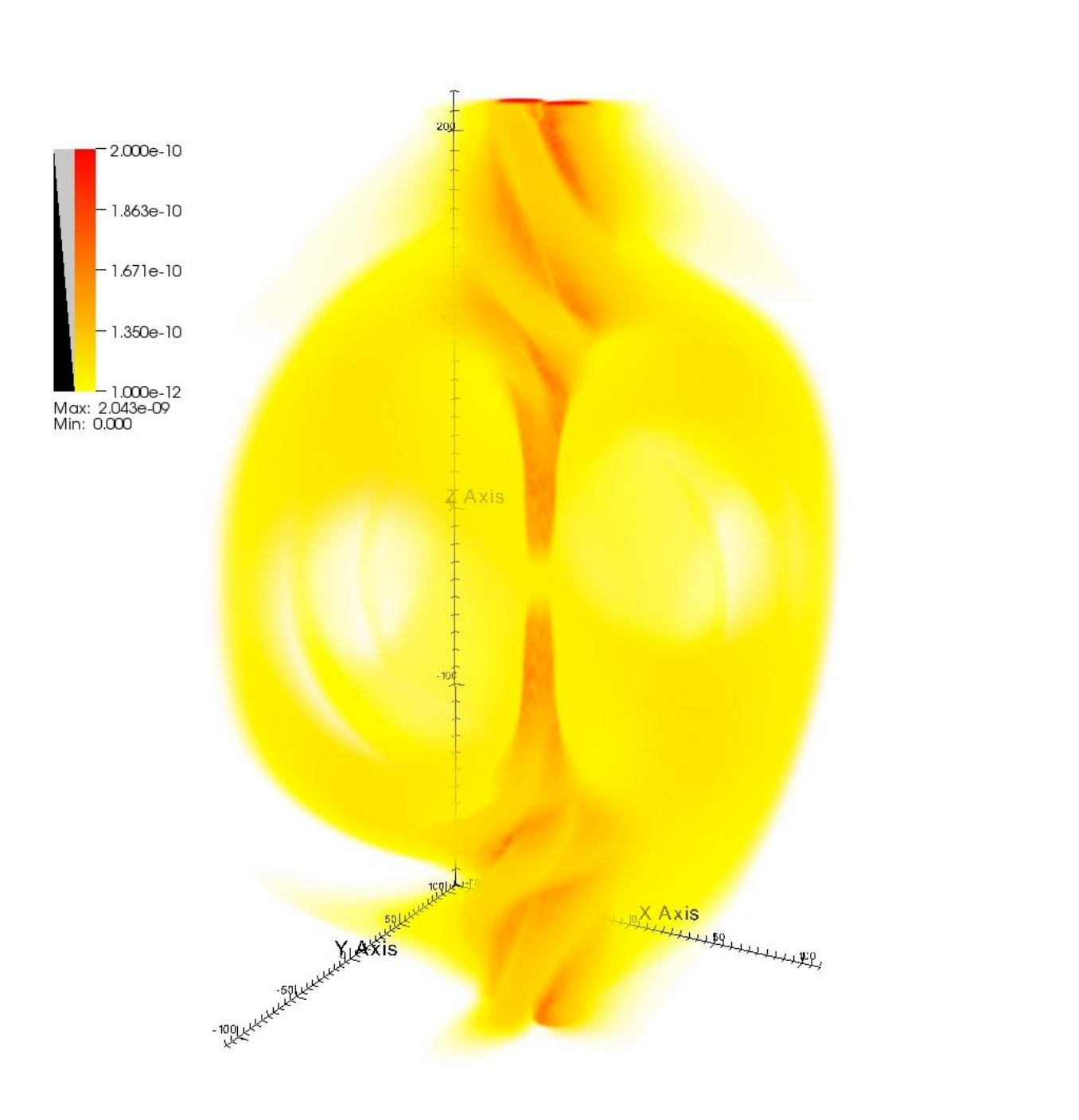} \label{fig:fourfigures-c}}
\subfloat[Part 4][$6.8\, M_8$ hrs]{\includegraphics[scale=0.25]{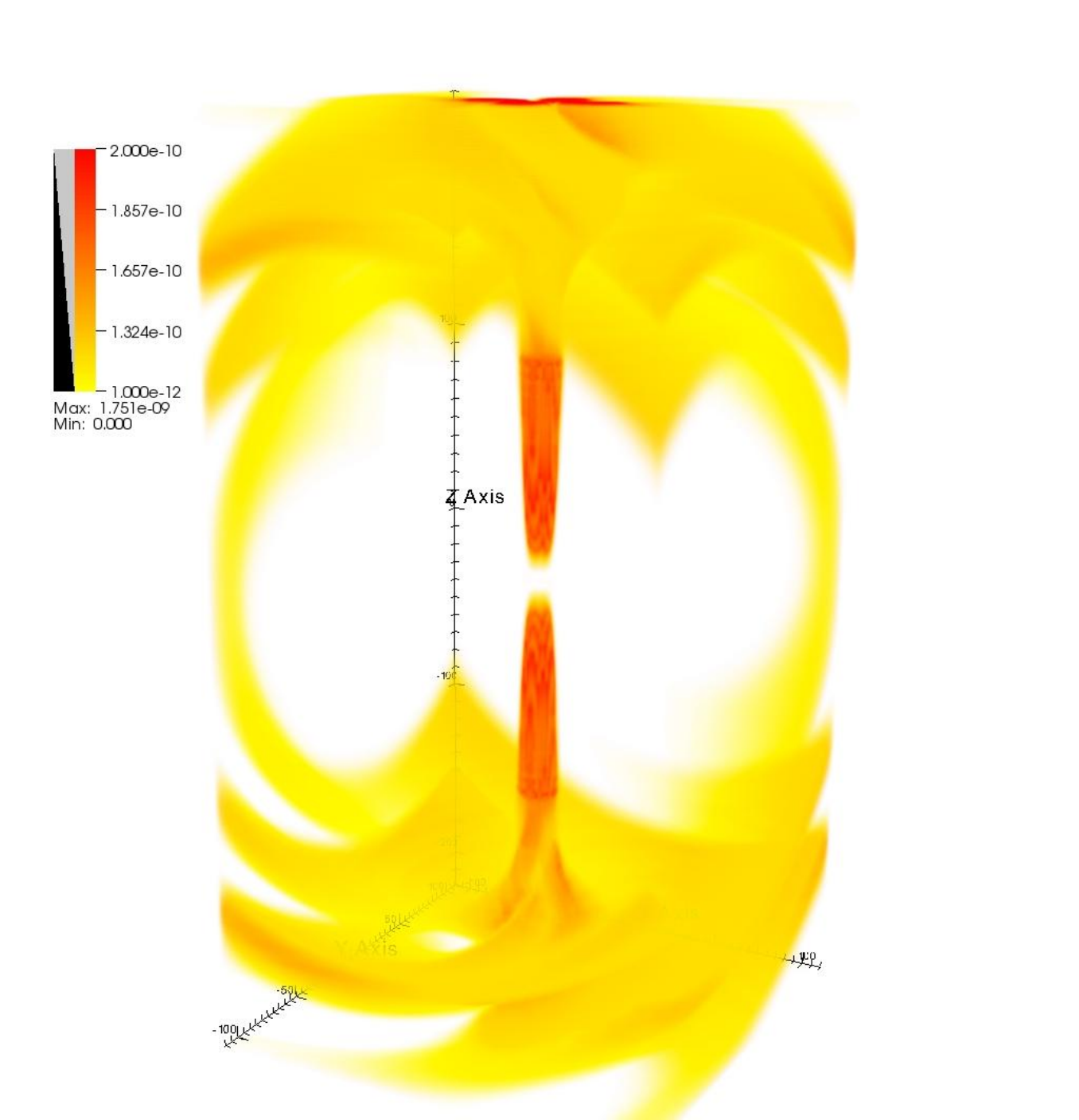} \label{fig:fourfigures-d}}
\caption{Electromagnetic energy flux at different times. The collimated
part is formed by two tubes orbiting around each other following the motion of the black holes.
A strong isotropic emission occurs at the time of merger, followed by a single
collimated tube as described by the Blandford-Znajek scenario.}
\label{fig:fourfigures}
\end{figure}


In the first stage, before the merger takes place, the black holes act as stirrers for the
surrounding plasma. Their orbital dynamics and strong curvature affect the electromagnetic field
in the close neighborhood of each black hole, inducing both a poloidal electric field
and a toroidal magnetic field --both scaling as $v B_0$-- (as in models for magnetospheric
interactions of binary pulsars~\cite{1996ApJ...471L..95V,2005ApJ...631..488R}). 
Fig. 2 illustrates field lines corresponding to $E$ and $B$
as well as the $\Omega_F$. The induced time-dependent topology gives rise to
a net Poynting flux aligned/antialigned with the orbital momentum vector 
around each black hole, with an EM frequency given by $\Omega_F \sim \Omega_{\rm orb}/5$. As a result,
despite the fact that the black holes are not spinning, there is a strongly collimated
electromagnetic flux of energy dominated by an $m=2$ multipolar structure with a time-dependence
determined by the orbital motion.

\begin{figure}
\centering
\subfloat[$-8.2 \, M_8$ hrs]{\includegraphics[width=1.1in,height=1.1in]{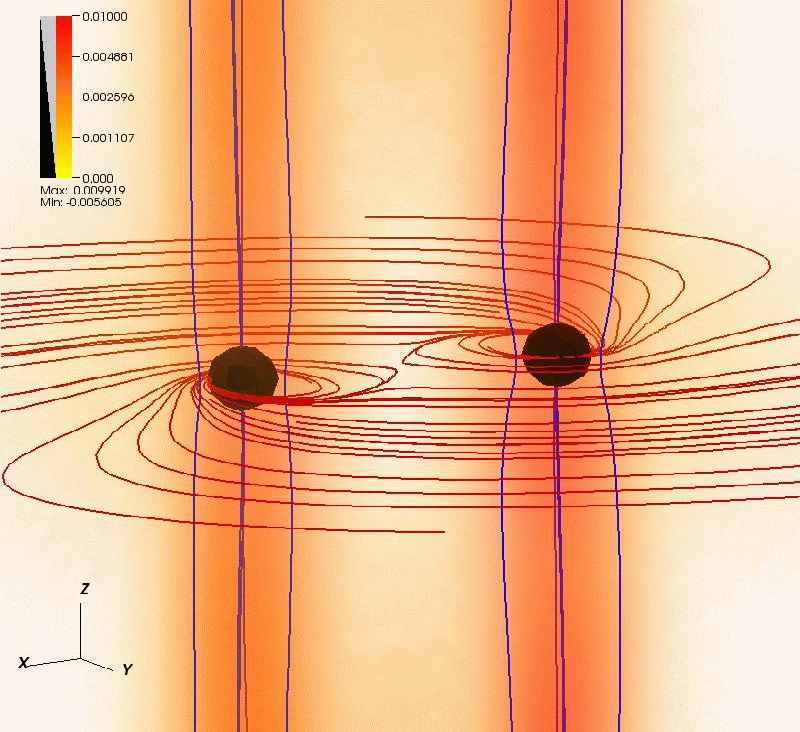}} \, \,
\subfloat[$4.6 \, M_8$ hrs]{\includegraphics[width=1.1in,height=1.1in]{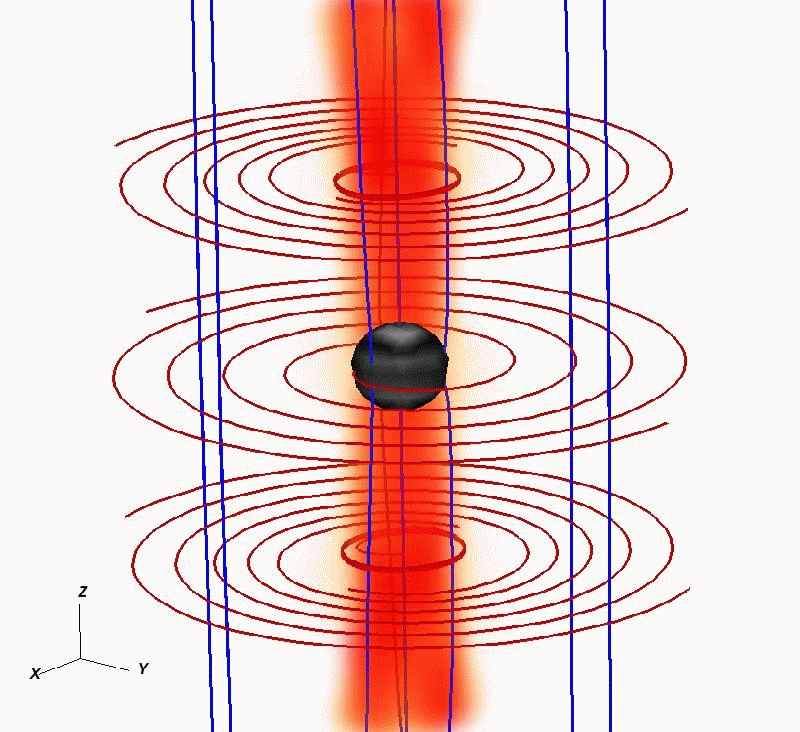}}
\caption{Some representative electric and magnetic field lines, together with the electromagnetic rotation
frequency $\Omega_F$.}
\end{figure}

During the merger, which lasts about 
$\sim 7 \, M_8$ hours
 --from late orbiting, through the plunge
phase to the formation of a highly distorted black hole--
the highly non-linear dynamics translates into a significant
increase in both electromagnetic and gravitational energy radiated. As expected, a common collimated 
tube arises and a natural transition from $m=2 \rightarrow 0$ is observed. In addition to the collimated
flux of energy, a rather isotropic flux is also emitted whose energy is much smaller than
the collimated one during the inspiral but of the same order at the merger, where
both these energy fluxes reach a peak and the collimated part doubles in magnitude.

Afterwards, the late stage is described by a single black hole which, after a relatively short time,
is well approximated by a Kerr black hole.
%
The remnant black hole  radiates gravitational radiation described by quasinormal modes which decay to zero
exponentially as the black hole settles into a Kerr configuration.
Thus, the post-merger behavior of the electromagnetic
field behavior is increasingly better represented by the Blandford-Znajek process for a spinning
black hole with $a \simeq 0.67$. As a result, the collimated electromagnetic energy flux does not decay to zero, rather it
approaches the value predicted by the Blandford-Znajek model. For a single spinning black hole,
this energy flux evaluated at the horizon goes like
$F_{EM} \sim R_{BH} \Omega_F (\Omega_H - \Omega_F) B^2$, where
$\Omega_H$ is the rotation frequency of the black hole (which is similar
to the orbital velocity at the merger). In this case, we numerically find that
the EM rotation frequency for a single black hole relaxes to $\Omega_F \sim \Omega_{\rm H}/2$.

Energetically, the system radiates gravitational waves primarily through 
$l=|m|= 2$ modes. These waves display a chirping behavior as the orbit tightens, followed by exponential
decay after the merger (see e.g.~\cite{Aylott:2009ya}). Overall,
the system radiates $\simeq 2.5\%$  and  $\simeq 16\%$ 
of the rest mass energy and angular momentum (at the initial separation) respectively. 
In the electromagnetic band the radiation
profile displays a more complex structure. 
Throughout the orbiting stages, the electromagnetic radiation
exhibits a strongly collimated character along the magnetic field lines in the region around
the individual black holes, together with a weaker isotropic emission. 
These features are illustrated in Fig.~1 in which the flux of electromagnetic energy is shown at four
different times during the evolution. 
In the early stages the black holes stir the surrounding plasma,
leading to a clear collimation of electromagnetic flux induced by the orbiting black holes pulling
the EM field lines. 
These collimated tubes twist about each other as the black holes
proceed. When the merger takes place, the collimated tubes merge into one and acquire
a rather smooth structure around the final black hole's ergosphere. In addition to the collimated
energy flux, a strong burst of isotropic electromagnetic radiation is produced at the merger.

Figure 3 illustrates the behavior of the electromagnetic energy flux over a sphere located at 
$r = 196 \,M_8 AU$. 
A clear collimation is observed which is evident in the ``bright-spots''. These spots are symmetric
with respect to the orbital plane and revolve around each other as a result of the orbiting until that
they merge into a single one along the poles. In addition to these
collimated flows, the figure also illustrates the energy flux through different latitudes and the
transition from the $m=2$ structure to $m=0$.

\begin{figure}
\centering
\subfloat[Part 1][$-8.2 \,M_8$ hrs]{\includegraphics[width=.9in,height=.9in]{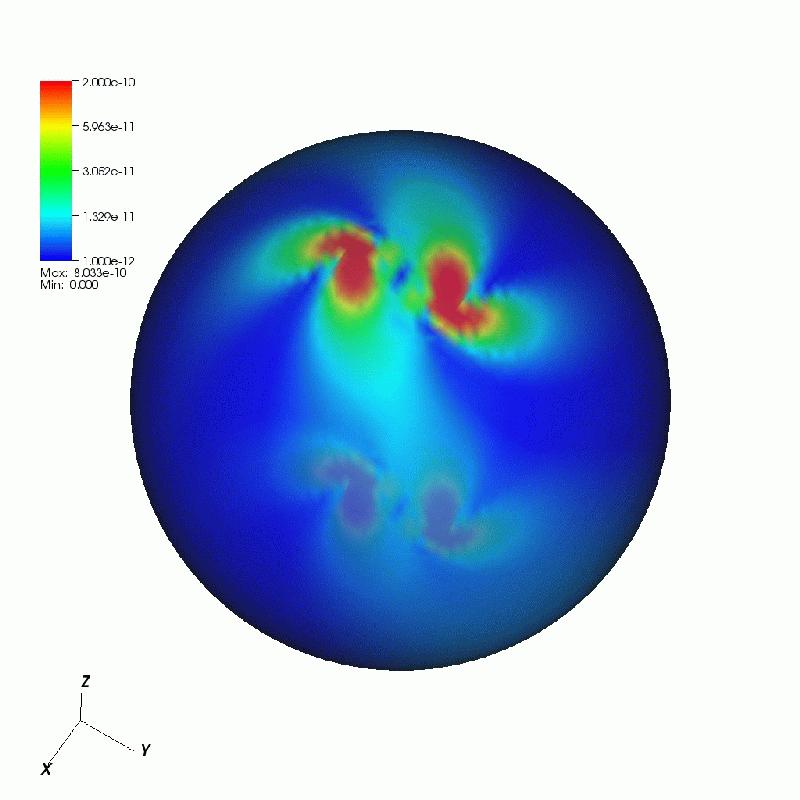}} \,\,\,\,\, 
\subfloat[Part 2][$-5.5 \,M_8$ hrs]{\includegraphics[width=.9in,height=.9in]{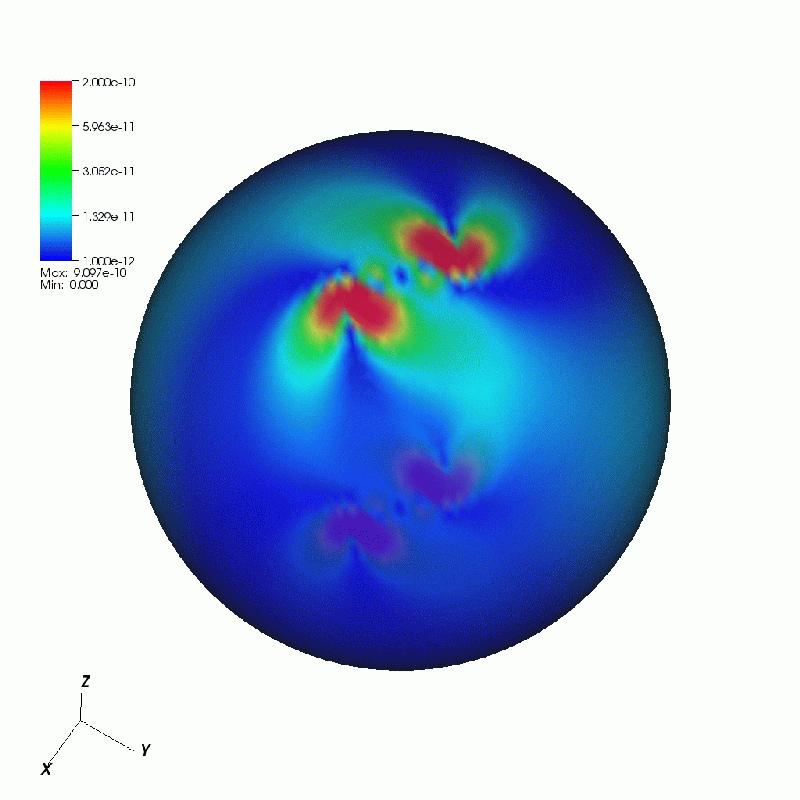}} \,\,\,\,\,
\subfloat[Part 3][$-3.0\, M_8$ hrs]{\includegraphics[width=.9in,height=.9in]{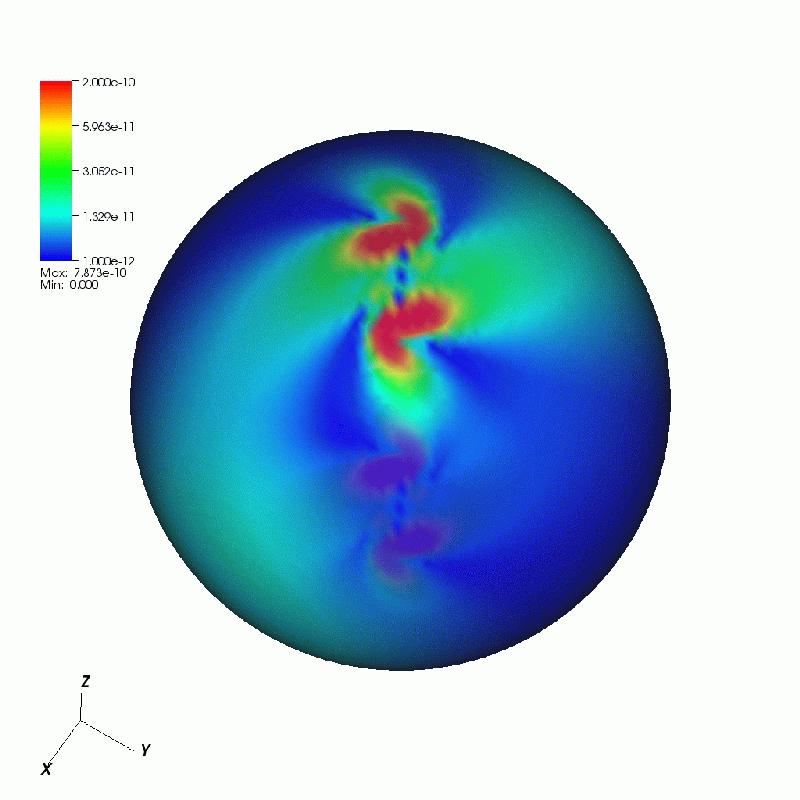}} \\
\subfloat[Part 4][$2.0 \,M_8$ hrs]{\includegraphics[width=.9in,height=.9in]{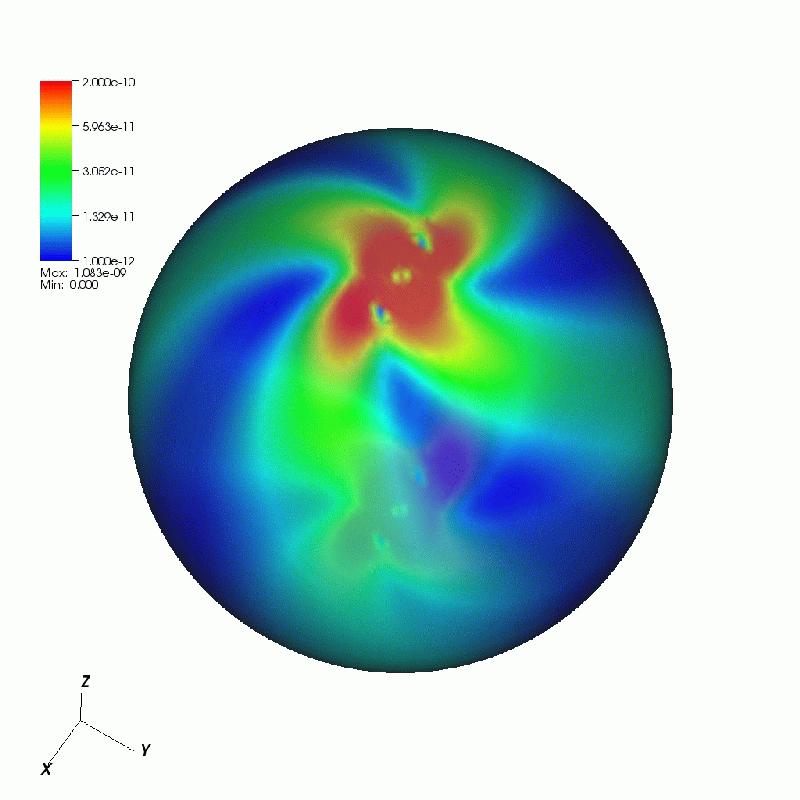}} \,\,\,\,\,
\subfloat[Part 5][$4.6 \,M_8$ hrs]{\includegraphics[width=.9in,height=.9in]{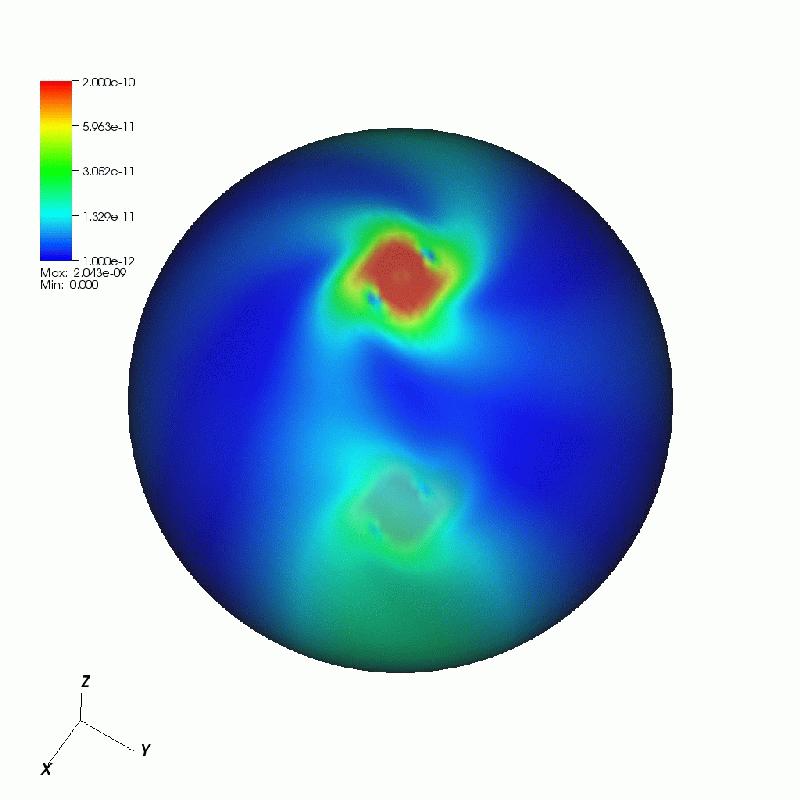}} \,\,\,\,\,
\subfloat[Part 6][$6.8 \,M_8$ hrs]{\includegraphics[width=.9in,height=.9in]{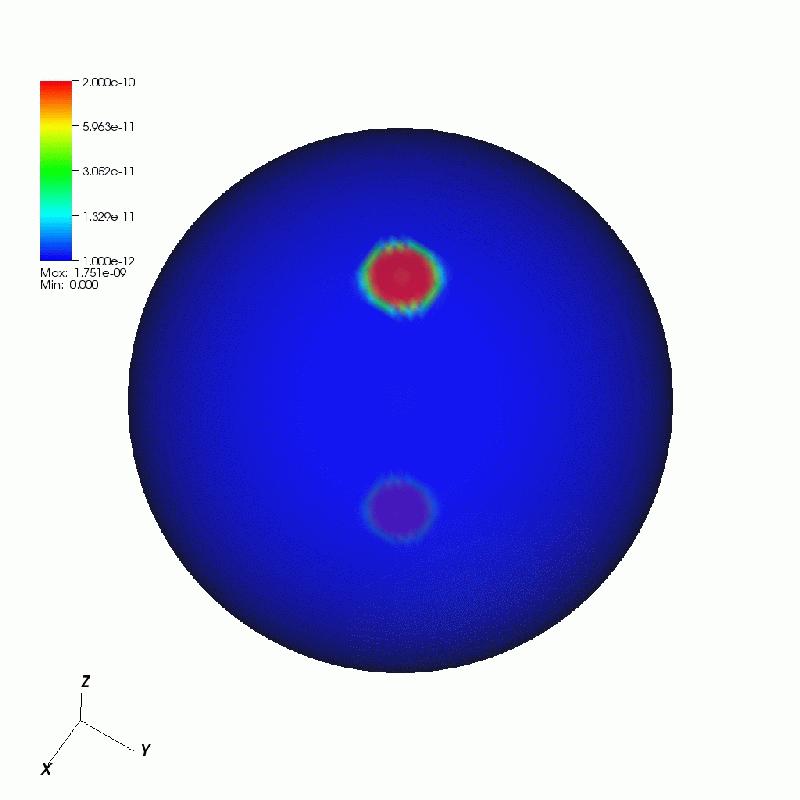}} 
\caption{Projection of the radial electromagnetic flux on a sphere located at $r = 20M$.
The structure of the tubes and the transition from a $m=2$ to a $m=0$ mode are clearly
displayed.}
\label{fig:fourfigures}
\end{figure}

Figure 4 provides the total energy flux in both the electromagnetic and gravitational wavebands
for a particular astrophysically relevant case with $M = 10^8\,M_{\odot}$ and $B = 10^{4}\,G$ 
(i.e. not exceeding the Eddington magnetic field strength $B\simeq 6 \cdot 10^{4}
M_8^{-1/2}\,G$~\cite{2008arXiv0810.1055D}). 

\begin{figure}
\centering
\subfloat{\includegraphics[width=1.6in,height=1.1in]{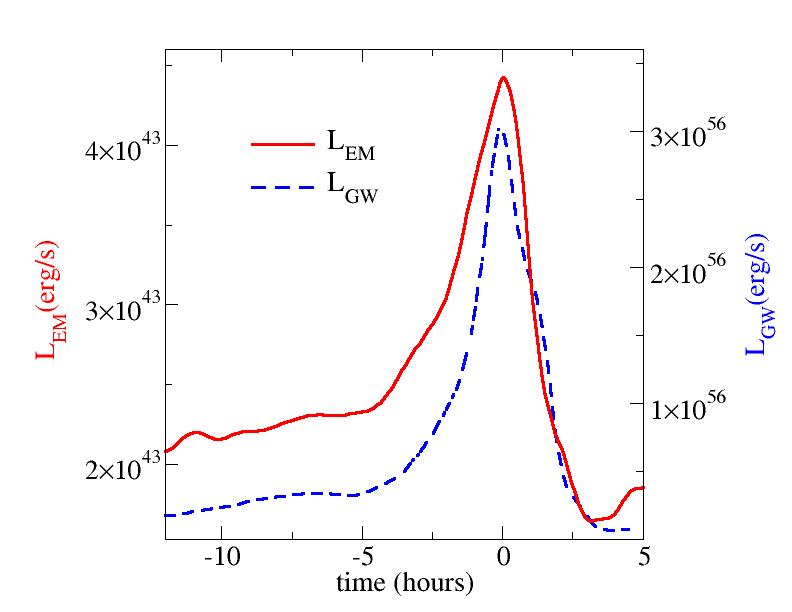}} \, \,
\subfloat{\includegraphics[width=1.4in,height=1.1in]{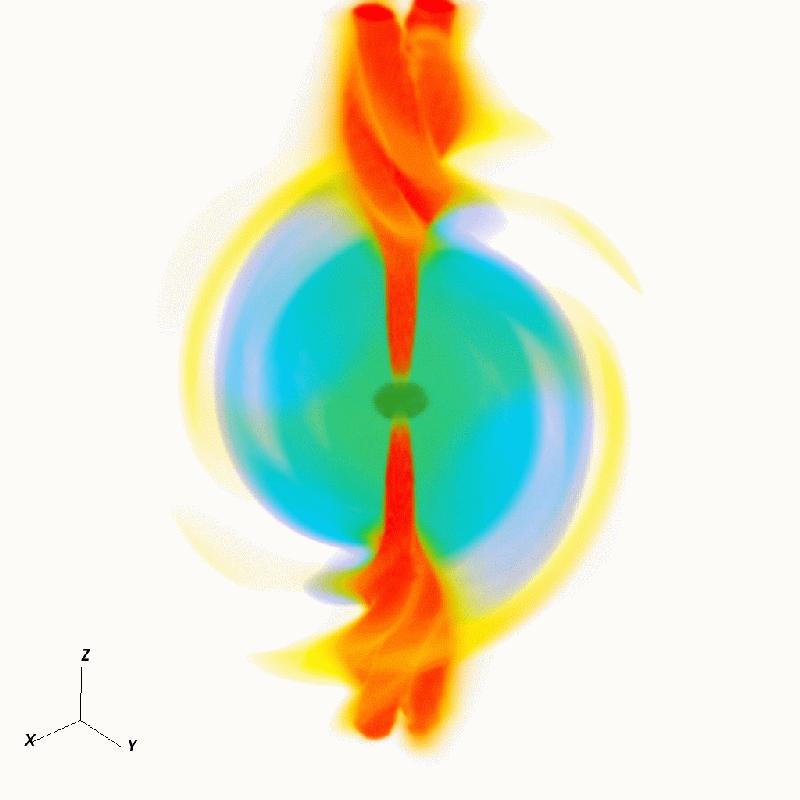}}
\caption{(Left) The gravitational and the (collimated) electromagnetic luminosities for the case $B_o=10^4\,G$
and $M=10^8 M_{\odot}$. The gravitational luminosity increases more than an order of magnitude
at the merger and then decays to zero. The electromagnetic luminosity during the inspiral
is only a factor $ \sim 2$ smaller than at the merger. After the peak, it decays to
a constant value given by the Blandford-Znajek mechanism for a single spinning black hole.
(Right) Electromagnetic (yellow-red) and gravitational (green-blue) radiation from the system
at $4.6 M_8$ hrs.}
\label{fig:fourfigures}
\end{figure}

A clear sweep upwards is apparent in both channels as the merger takes place. Afterwards,
both decrease rapidly. However it is interesting to note that while the gravitational flux essentially shuts off in a short time (as
the system relaxes to a stationary black hole), the electromagnetic flux rises up to the level
predicted by the Blandford-Znajek mechanism.

Overall, the total (Poynting flux) radiated electromagnetic energy behaves like
$E^{em}_{rad}/(M c^2) = 10^{-14}\, \left( M_8 \right)^2 \, \left( B_4 \right)^2$ (with 
$B_4 \equiv B/(10^4G)$.
This is about an order of magnitude larger than the same system studied
in the electrovacuum case (i.e. not including the plasma) \cite{Palenzuela:2009yr,Palenzuela:2009hx}, 
indicating that the plasma taps the orbital/rotational energy from the system more efficiently.
As studied in \cite{MPLRPY10}, the isotropic radiation transport of magnetic field perturbations
can be transmitted to the disk in the form of magnetosonic waves when the ratio between the viscous
time scales of the disk and the transport magnetic timescale is above $\simeq 1$. 
Thus, in~\cite{MPLRPY10} it was indicated that an electromagnetic counterpart to a
gravitational wave event could
be determined by a variability induced in a possible accretion mechanism taking place.

The results highlighted in the current work indicate that 
a more important observational prospect
is achieved by the jets induced by the binary's dynamics. These jets have
a (Poynting flux) luminosity during the last hours before the merger of 
$2-4\cdot10^{43} \,erg/s \sim 0.0015-0.0030\, L_{Edd}$  (with $L_{Edd}$ the Eddington luminosity).
This electromagnetic energy can be transferred to kinetic energy of the plasma, which will
radiate through synchrotron processes in frequencies around the GHz region. Such a flux of energy implies it would
be possible to observe these systems to $z \simeq 1$, with missions such as {\em IXO} and {\em EXIST}, depending
on the efficiency of the photon emission process.
As Fig 4. illustrates, the system will be sufficiently bright for hours before the merger and remain 
so through merger.
This radiated power has a time dependence given by $(R_{\rm orb} \, \Omega_{\rm orb})^2 B^2$ and the flux exhibits a clear
transition $m=2 \rightarrow 0$ at the time of the merger. Thus,
prospects for detecting pre-merger and prompt electromagnetic counterparts are certainly interesting. 
Furthermore, joint detections in both electric and
gravitational wave bands (the power of the latter scales as $R_{\rm orb}^4 \, \Omega_{\rm orb}^6 M^2$) 
are therefore quite probable as
LISA will be capable of observing supermassive binary black hole systems for weeks to months before
merger and considerable time afterwards.

\noindent{\bf{\em Conclusions:}}
Observational prospects of detecting gravitational waves with LISA from supermassive black hole
mergers will be excellent~\cite{Holz:2005df}. Indeed, LISA's superb noise spectra will allow for clean
determination of gravitational waves and extraction of key physical parameters of the
system such as certain combinations of the their masses and spins, orientation, position in the sky and 
luminosity distance. Gravitational waves from supermassive binary systems 
could be detected as far as redshifts of $z=5-10$~\cite{Vecchio:2003tn}. However,
LISA's localization in the sky--an ellipse with a few arcminutes to a few degrees in size- 
will not be sufficient to unequivocally determine the source's position~\cite{Holz:2005df,Lang:2008gh}, and 
it is
here where an electromagnetic
counterpart would aid tremendously in determining the location. Precise localization and confrontation of
the observed signals will allow
for tremendous physics pay off: general relativity, cosmology, astrophysics and even 
efforts towards a quantum theory of gravity will likely enjoy a revolution in their understanding~\cite{Holz:2005df,Bloom:2009vx,Haiman:2008zy}.

As we have shown here, binary black hole mergers could give rise to strong electromagnetic output with
(Poynting flux) luminosities as high as a few $10^{43}$ ergs/s. This would correspond to an isotropic bolometric 
flux of $F_x \simeq 10^{-15}$ erg/($cm^2$ s). Such a flux could be detected to a redshift of
$z=1$  and {\em possibly larger if accounting for anisotropies on the one hand, but on the other hand depending on the
photon emission process efficiency on the other}. Interestingly, our work indicates that
a marked transition in the electromagnetic flux from $m=2 \rightarrow 0$ will take place around the
merger time which will leave its imprint in possible observable signals.  
Furthermore, we have shown that even non-spinning black holes could give rise to
interesting levels of output, which becomes even stronger as the merger takes place.
Certainly, higher power would be expected 
if any of the individual black holes is spinning and/or the magnetic field strength
is larger in the region of the black holes either as sourced by the disk or further enhanced
by the merger process. In the latter case, a tantalizing option would be for such enhancement to be 
driven by an analog of the BZ effect; i.e. by significantly tapping
orbital rotational energy before the merger takes place as in the single black hole
case where rotational energy from the black hole is extracted.
 A requirement for this scenario would be 
that the black holes  orbit within an ergo-region
and before the plunge so that there is sufficient time to efficiently tap into this source of energy. 
As argued in~\cite{Palenzuela:2009hx}, this scenario appears unlikely 
unless the black holes are significantly spinning and aligned with the orbital angular momentum;
otherwise, the black holes would be in the plunging phase when crossing (an estimate of) the ergosphere.
While the magnitude of spins in supermassive black holes is still largely unknown --gravitational wave
detections will be the ultimate tool to understand this issue--
alignment of spins with the orbital angular momentum is expected in a large class of systems~\cite{Kesden:2010ji}.

\noindent{\bf{\em Acknowledgments:}}
It is a pleasure to thank  J. Aarons, N. Afshordi, A. Broderick, P. Chang, T. Garrett,
B. MacNamara,  E. Poisson, A. Spitkovsky and C. Thompson as well as 
our long time collaborators M. Anderson, E. Hirschmann, P. Motl, M. Megevand and
D. Neilsen
for useful discussions and comments. We acknowledge 
support comes NSF grant PHY-0803629 to Louisiana State University and 
PHY-0803624 to Long Island University as well as NSERC through a Discovery Grant.  
Research at Perimeter Institute is
supported through Industry Canada and by the Province of Ontario
through the Ministry of Research \& Innovation.  Computations were
performed Teragrid and Scinet.
CP and LL thank the Princeton Center for Theoretical Physics for hospitality
where parts of this work were completed.

\bibliographystyle{apsrev}

\end{document}